A Torsional Topological Invariant

(To appear in *Proceedings for Conference in Honor of CN Yang's 85th Birthday*)

H. T. Nieh

Center for Advanced Study, Tsinghua University, Beijing 100084, China

and

C.N. Yang Institute for Theoretical Physics, State University of New York at Stony Brook

Stony Brook, New York 11794, U.S.A.

Abstract

Curvature and torsion are the two tensors characterizing a general Riemannian space-time. In Einstein's General Theory of Gravitation, with torsion postulated to vanish and the affine connection identified to the Christoffel symbol, only the curvature tensor plays the central role. For such a purely metric geometry, two well-known topological invariants, namely the Euler class and the Pontryagin class, are useful in characterizing the topological properties of the space-time. From a gauge theory point of view, and especially in the presence of spin, torsion naturally comes into play, and the underlying spacetime is no longer purely metric. We describe a torsional topological invariant, discovered in 1982, that has now found increasing usefulness in recent developments.

Professor C. N. Yang played a leading role in laying the foundation for the development of gauge theories in particle physics. His hallmark contribution is the Yang-Mills nonabelian gauge theory [1]. My first learning of the Yang-Mills gauge theory came in 1963 when Professor Julian Schwinger, my thesis advisor, assigned me to investigate the question of mass of the Yang-Mills gauge particle; apparently he himself was thinking about the mass problem of gauge bosons during that period [2]. After struggling for about a year, of course, nothing came out of it. It was also around that time Schwinger taught in his quantum field theory class Herman Weyl's 1929 formulation of Dirac electrons in the gravitational field [3], in which Einstein's gravitational theory is cast as a gauge theory of local Lorentz gauge symmetry. In the mid 1970's, after the standard model of Glashow-Weinberg-Salam had more or less settled down and after the appearance of Professor Yang's 1974 paper on gravitation [4], I became interested in learning again about the theory of gravitation. In the spirit of gauge theory, in which the connection, or the gauge field, plays central role, Weyl's and Fock's pioneering formulation [3] together with its revival by Kibble [5] and Sciama [6] in the early 1960's clearly indicate that torsion [7], in addition to curvature, could play an important role in the development of gravitational theory. It was during this learning process that an identity relating the totally antisymmetric part of the curvature tensor with torsion was found, which M.L. Yan and I later published in 1982 [8]. This leads to a topological invariant that characterizes the torsional property of the spacetime, and has found increasing usefulness in recent developments. It is this work I will report on .

## Riemann-Cartan Space-time

The Yang-Mills gauge theory is based on the premise of local freedom in defining isospin. To make "compatible" the definitions of isopin at neighboring spacetime points, a connection field, or the Yang-Mills gauge field, is introduced. Einstein's theory of general relativity is based on Riemannian geometry, in which local definitions of vectors and tensors at neighboring space-time points are correlated by the familiar Christoffel connection. The Christoffel connection, though expressed in terms of the metric tensor, thus plays the role of a gauge field for the group of general coordinate transformations GL(4). Other than electromagnetic theory, the theory of general relativity is indeed the earliest gauge theory, and with a very rich structure. The Dirac spinor, on the other hand, is a two-valued representation of the Lorentz group SO(3,1), which has very different transformation properties from vector representations. To accommodate local freedom of defining spinor property with respect to local Lorentz frames at neighboring spacetime points, Weyl and Fock [3] introduced a new connection, the Lorentz connection or spin-connection, to relate local definitions of a Dirac spinor. Again, this is very much a play of the gauge concept. Under Lorentz transformations of the local Cartesian frame, which is represented by the four orthonormal vectors $e^a{}_\mu(x)$, the "vierbeins", the Dirac spinor $\Psi(x)$ transforms according to [9]

$$\Psi(x) \to e^{-i\varepsilon^{ab}(x)\sigma_{ab}/4}\Psi(x) \tag{1}$$

where

$$\sigma_{ab} = \frac{i}{2}[\gamma_a, \gamma_b],$$

$$\{\gamma_a, \gamma_b\} = 2\eta_{ab},$$

$$\eta_{ab} = (1,-1,-1,-1).$$

The Roman letters a, b, etc. are the frame indices, while the Greek letters $\mu$, $\nu$, *etc*. the coordinate indices. The corresponding connection field $\omega^{ab}{}_\mu(x)$, commonly called the spin connection or Lorenz connection, is introduced such that the covariant derivative

$$D_\mu \Psi \equiv (\partial_\mu - \frac{i}{4}\omega^{ab}{}_\mu \sigma_{ab})\Psi \tag{2}$$

transforms in a covariant way:

$$D_\mu \Psi(x) \to e^{-i\varepsilon^{ab}(x)\sigma_{ab}/4}\Psi(x). \tag{3}$$

This requires that the spin connection field $\omega^{ab}{}_\mu(x)$ transforms according to

$$\omega_\mu(x) \to \omega'_\mu(x) = e^{-i\varepsilon(x)} \omega_\mu(x) e^{i\varepsilon(x)} - [i\partial_\mu e^{-i\varepsilon(x)}] e^{i\varepsilon(x)}, \quad (4)$$

where

$$\omega_\mu(x) \equiv \frac{1}{4} \omega^{ab}{}_\mu(x) \sigma_{ab},$$

$$\varepsilon(x) \equiv \frac{1}{4} \varepsilon^{ab}(x) \sigma_{ab}.$$

The spacetime metric is defined by

$$g_{\mu\nu} = \eta_{ab} e^a{}_\mu e^b{}_\nu \quad (5)$$

while the proper definition for the GL(4) connection, which plays the role of the gauge field for general coordinate transformations, is given by [5]

$$\Gamma^\lambda{}_{\mu\nu} \equiv e_a{}^\lambda (e^a{}_{\mu,\nu} + \omega^a{}_{b\nu} e^b{}_\mu). \quad (6)$$

The "minimum" combination

$$\Psi i\gamma^a e_a{}^\mu (\partial_\mu - \frac{i}{4} \omega^{bc}{}_\mu \sigma_{bc}) \Psi + h.c., \quad (7)$$

where $e_a{}^\mu$ is the inverse of $e^a{}_\mu$, being invariant under local Lorentz transformations as well as under the GL(4) general coordinate transformations when the transformation properties of the vierbein fields are correspondingly defined, is the natural choice as the Dirac Lagrangian in curved spacetime.

We note that, in this formulation, there are two sets of field variables: the spin-connection fields $\omega^{ab}{}_\mu(x)$ and the "vierbein" fields $e^a{}_\mu$. In defining covariant derivatives, $\omega^{ab}{}_\mu(x)$ and $\Gamma^\lambda{}_{\mu\nu}$ are the gauge fields for the local Lorentz transformations and the general coordinate transformations, respectively. The corresponding "curvature" tensors or field strengths are given, respectively, by

$$R^{ab}{}_{\mu\nu} \equiv \omega^{ab}{}_{\mu,\nu} - \omega^{ab}{}_{\nu,\mu} - \omega^{ac}{}_\mu \omega_c{}^b{}_\nu + \omega^{ac}{}_\nu \omega_c{}^b{}_\mu, \quad (8)$$

$$R^{\lambda\rho}{}_{\mu\nu} \equiv g^{\rho\sigma} (\Gamma^\lambda{}_{\sigma\mu,\nu} - \Gamma^\lambda{}_{\sigma\nu,\mu} - \Gamma^\lambda{}_{\alpha\mu} \Gamma^\alpha{}_{\sigma\nu} + \Gamma^\lambda{}_{\alpha\nu} \Gamma^\alpha{}_{\sigma\mu}). \quad (9)$$

Having the property:

$$R^{\lambda\rho}{}_{\mu\nu} = e_a{}^\lambda e_b{}^\rho R^{ab}{}_{\mu\nu}, \quad (10)$$

these two curvature tensors are closely related.

There are the following two possibilities, yielding different theories [10]: (i) Both $\omega^{ab}{}_\mu(x)$

and $e^a{}_\mu$ are taken to be independent field variables in the theory and are to be determined by the theory. (ii) $\omega^{ab}{}_\mu(x)$ is given by the Ricci coefficients of rotation in terms of $e^a{}_\mu$ and has the property of satisfying the required transformation property [5]:

$$\omega_{ab\mu} = \frac{1}{2} e^c{}_\mu (\gamma_{cab} - \gamma_{abc} - \gamma_{bca}) \tag{11}$$

where

$$\gamma^c{}_{ab} = (e_a{}^\mu e_b{}^\nu - e_b{}^\mu e_a{}^\nu) e^c{}_{\mu,\nu}$$

When one opts for the possibility (ii), only $e^a{}_\mu$ are basic field variables of the theory and the GL(4) connection field $\Gamma^\lambda{}_{\mu\nu}$ defined by (7) is the Christoffel connection:

$$\Gamma^\lambda{}_{\mu\nu} = \frac{1}{2} g^{\lambda\rho} (g_{\rho\mu,\nu} + g_{\nu\rho,\mu} - g_{\mu\nu,\rho}), \tag{12}$$

which is symmetric and yields vanishing torsion tensor $C^\lambda{}_{\mu\nu}$, where the torsion tensor is defined by

$$C^\lambda{}_{\mu\nu} \equiv \Gamma^\lambda{}_{\mu\nu} - \Gamma^\lambda{}_{\nu\mu} = e_a{}^\lambda (e^a{}_{\mu,\nu} - e^a{}_{\nu,\mu} + \omega^a{}_{b\nu} e^b{}_\mu - \omega^a{}_{b\mu} e^b{}_\nu). \tag{13}$$

If, on the other hand, we accept spinors as physically fundamental and regard Lorentz group as the fundamental gauge group, then the spin connection $\omega^{ab}{}_\mu(x)$ should be regarded as basic field variables. Namely, one would opt for the possibility (i) mentioned above. The GL(4) connection field $\Gamma^\lambda{}_{\mu\nu}$ as defined by (7) is then, in general, not symmetric: $\Gamma^\lambda{}_{\mu\nu} \neq \Gamma^\lambda{}_{\nu\mu}$, giving rise to non-vanishing torsion tensor. It is clear from the Dirac Lagrangian (5) that *when the connection field or the gauge field $\omega^{ab}{}_\mu(x)$ is taken to be an independent variable in the theory*, it receives contribution from the Dirac field in the form of

$$\frac{1}{4} e^{a\mu} \Psi (\gamma_a \sigma_{bc} + \sigma_{bc} \gamma_a) \Psi,$$

or,

$$\frac{1}{2} \eta_{abcd} e^{a\mu} \Psi \gamma_5 \gamma^d \Psi. \tag{14}$$

This is the origin of a non-vanishing torsion tensor. It is thus seen that torsion could play an important role in gauge theories of gravitation when the Dirac field is brought into the system of consideration.

## Gauss-Bonnet Identities

In purely metric Riemannian space-time, the Euler and Pontryagin 4-forms

$$\sqrt{-g}\,\varepsilon_{\alpha\beta\gamma\delta}\varepsilon^{\mu\nu\lambda\rho}R^{\alpha\beta}{}_{\mu\nu}R^{\lambda\delta}{}_{\lambda\rho}\,, \quad \text{(Euler)}$$

$$\sqrt{-g}\,\varepsilon^{\mu\nu\lambda\rho}R^{\alpha\beta}{}_{\mu\nu}R_{\alpha\beta\lambda\rho}\,, \quad \text{(Pontryagin)}$$

are well known to satisfy the Gauss-Bonnet identities. In the case of non-vanishing torsion, the identification (10) allows verification of these identities [11] by making use of the Clifford algebra satisfied by the Dirac matrices.

Define

$$\varpi_\mu \equiv \frac{1}{4}\omega^{ab}{}_\mu \sigma_{ab}, \tag{15}$$

$$\overline{R}_{\mu\nu} \equiv \varpi_{\mu,\nu} - \varpi_{\nu,\mu} - i[\varpi_\mu,\varpi_\nu] = \frac{1}{4}R^{ab}{}_{\mu\nu}\sigma_{ab}, \tag{16}$$

where in the last step use has been made of the Lorentz algebra:

$$\frac{i}{2}[\sigma_{ab},\sigma_{cd}] = \eta_{ac}\sigma_{bd} - \eta_{ad}\sigma_{bc} + \eta_{bd}\sigma_{ac} - \eta_{bc}\sigma_{ad}. \tag{17}$$

Denoting by $\eta_{abcd}$ the totally anti-symmetric Minkowski tensor, with $\eta_{0123} = -1$, and noticing

$$\eta_{abcd}e^a{}_\mu e^b{}_\nu e^c{}_\lambda e^d{}_\rho = \varepsilon_{\mu\nu\lambda\rho}, \quad (\det e^a{}_\mu)^2 = -\det g_{\mu\nu} = -g\,, \tag{18}$$

the Euler 4-form can be expressed in the form:

$$\begin{aligned}
&\sqrt{-g}\,\varepsilon_{\alpha\beta\gamma\delta}\varepsilon^{\mu\nu\lambda\rho}R^{\alpha\beta}{}_{\mu\nu}R^{\lambda\delta}{}_{\lambda\rho}\\
&= \sqrt{-g}\,\eta_{abcd}\varepsilon^{\mu\nu\lambda\rho}R^{ab}{}_{\mu\nu}R^{cd}{}_{\lambda\rho}\\
&= 4i\sqrt{-g}\,\varepsilon^{\mu\nu\lambda\rho}Tr\left[\gamma_5 \overline{R}_{\mu\nu}\overline{R}_{\lambda\rho}\right]
\end{aligned} \tag{19}$$

where, with

$$\gamma_5 = i\gamma^0\gamma^1\gamma^2\gamma^3,$$

use has been made of

$$Tr[\gamma_5 \sigma_{ab}\sigma_{cd}] = -4i\eta_{abcd}.$$

On account of (14) and

$$\varepsilon^{\mu\nu\lambda\rho}Tr[\gamma_5 \varpi_\mu \varpi_\nu \varpi_\lambda \varpi_\rho] = 0,$$

$$[\gamma_5,\varpi_\mu] = 0,$$

(17) becomes

$$\sqrt{-g}\,\varepsilon_{\alpha\beta\gamma\delta}\varepsilon^{\mu\nu\lambda\rho}R^{\alpha\beta}{}_{\mu\nu}R^{\lambda\delta}{}_{\lambda\rho}$$

$$= 16i\sqrt{-g}\,\varepsilon^{\mu\nu\lambda\rho}Tr\{\lambda_5[\varpi_{\mu,\nu}\varpi_{\lambda,\rho} - i\varpi_{\mu,\nu}\varpi_\lambda\varpi_\rho - i\varpi_\mu\varpi_\nu\varpi_{\lambda,\rho}]\}$$

$$= 16i\sqrt{-g}\,\varepsilon^{\mu\nu\lambda\rho}Tr\{\gamma_5[(\partial_\mu\varpi_\nu)(\partial_\lambda\varpi_\rho) + \frac{2i}{3}\partial_\mu(\varpi_\nu\varpi_\lambda\varpi_\rho)]\} \quad (20)$$

$$= \partial_\mu\{16i\sqrt{-g}\,\varepsilon^{\mu\nu\lambda\rho}Tr[\gamma_5(\varpi_\nu\partial_\lambda\varpi_\rho + \frac{2i}{3}\varpi_\nu\varpi_\lambda\varpi_\rho)]\}.$$

This is the Gauss-Bonnet formula for the Euler 4-form. It is due to this property of being a total derivative that the volume integral of the Euler 4-form is a topological invariant.

Along the same vein, we can verify that in the case of non-vanishing torsion the Pontryagin 4-form satisfies the following Gauss-Bonnet type identity:

$$\sqrt{-g}\,\varepsilon^{\mu\nu\lambda\rho}R^{\alpha\beta}{}_{\mu\nu}R_{\alpha\beta\lambda\rho} = 2\sqrt{-g}\,\varepsilon^{\mu\nu\lambda\rho}Tr[\overline{R}_{\mu\nu}\overline{R}_{\lambda\rho}]$$

$$= \partial_\mu[8\sqrt{-g}\,\varepsilon^{\mu\nu\lambda\rho}Tr(\varpi_\nu\partial_\lambda\varpi_\rho + \frac{2i}{3}\varpi_\nu\varpi_\lambda\varpi_\rho)]. \quad (21)$$

We can extend the above derivation to a larger set of field variables. The larger set [8,11] contains the spin connection fields $\omega^{ab}{}_\mu(x)$ and the vierbein fields $e^a{}_\mu$, which we group together to form antisymmetric $\omega^{AB}{}_\mu$ (A,B = 0,1,2,3,5)

$$\omega^{AB}{}_\mu = \omega^{ab}{}_\mu \quad \text{(for A,B = 0,1,2,3)} \quad (22)$$

$$\omega^{a5}{}_\mu = \frac{1}{l}e^a{}_\mu, \text{ (a = 0,1,2,3)} \quad (23)$$

where a constant $l$ with the dimension of length is added to match the dimension of $\omega^{ab}{}_\mu(x)$ [12].

As $\gamma^0, \gamma^1, \gamma^2, \gamma^3, \gamma^5$ form a set five anti-commuting matrices, we can easily construct a set of anti-symmetric $X_{AB}$ (A, B = 0,1,2,3,5) satisfying the de Sitter or anti-de Sitter algebra:

$$\frac{i}{2}[X_{AB}, X_{CD}] = \eta_{AC}X_{BD} - \eta_{AD}X_{BC} + \eta_{BD}X_{AC} - \eta_{BC}X_{AD}, \quad (24)$$

with $\eta_{AB} = (1,-1,-1,-1,\pm 1)$. Letting

$$\Omega_\mu \equiv \frac{1}{4}X_{AB}\varpi^{AB}{}_\mu, \quad (25)$$

we define $\overline{F}_{\mu\nu}$ and $F^{AB}{}_{\mu\nu}$ according to

$$\overline{F}_{\mu\nu} \equiv \Omega_{\mu,\nu} - \Omega_{\nu,\mu} - i[\Omega_\mu, \Omega_\nu] = \frac{1}{4}X_{AB}F^{AB}{}_{\mu\nu}. \quad (26)$$

The $F^{AB}{}_{\mu\nu}$ so defined is given by, according to (24),

$$F^{AB}{}_{\mu\nu} = \omega^{AB}{}_{\mu,\nu} - \omega^{AB}{}_{\nu,\mu} + \eta_{CD}(\omega^{AC}{}_\mu \omega^{BD}{}_\nu - \omega^{AC}{}_\nu \omega^{BD}{}_\mu), \tag{27}$$

which has the contents (a,b = 0,1,2,3) :

$$F^{ab}{}_{\mu\nu} = R^{ab}{}_{\mu\nu} + \frac{1}{l^2}\eta_{55}(e^a{}_\mu e^b{}_\nu - e^a{}_\nu e^b{}_\mu), \tag{28}$$

$$F^{a5}{}_{\mu\nu} = \frac{1}{l}e^a{}_\lambda C^\lambda{}_{\mu\nu}, \tag{29}$$

where $C^\lambda{}_{\mu\nu}$ is the torsion tensor defined in (13). Following the same procedure as in deriving the identities (20) and (21), we can derive a similar identity for $F^{AB}{}_{\mu\nu}$:

$$\sqrt{-g}\,\varepsilon^{\mu\nu\lambda\rho} F^{AB}{}_{\mu\nu} F_{AB\lambda\rho} = 2\sqrt{-g}\,\varepsilon^{\mu\nu\lambda\rho} Tr[\overline{F}_{\mu\nu}\overline{F}_{\lambda\rho}]$$
$$= \partial_\mu[8\sqrt{-g}\,\varepsilon^{\mu\nu\lambda\rho} Tr(\Omega_\nu \partial_\lambda \Omega_\rho + \frac{2i}{3}\Omega_\nu \Omega_\lambda \Omega_\rho)]. \tag{30}$$

Torsional Topological Invariant

On account of (28) and (29), we have the difference of the two Pontryagin 4-forms:

$$\sqrt{-g}\,\varepsilon^{\mu\nu\lambda\rho} F^{AB}{}_{\mu\nu} F_{AB\lambda\rho} - \sqrt{-g}\,\varepsilon^{\mu\nu\lambda\rho} \overline{R}^{ab}{}_{\mu\nu} \overline{R}_{ab\mu\nu}$$
$$= -4\eta_{55}\frac{1}{l^2}\sqrt{-g}\,\varepsilon^{\mu\nu\lambda\rho}(R_{\mu\nu\lambda\rho} + \frac{1}{2}C^\lambda{}_{\mu\nu} C_{\lambda\mu\nu}). \tag{31}$$

Substracting (30) from (21) yields

$$\sqrt{-g}\,\varepsilon^{\mu\nu\lambda\rho}(R_{\mu\nu\lambda\rho} + \frac{1}{2}C^\alpha{}_{\mu\nu} C_{\alpha\lambda\rho}) = \partial_\mu(-\sqrt{-g}\,\varepsilon^{\mu\nu\lambda\rho} C_{\nu\lambda\rho}), \tag{32}$$

or, in the more compact form:

$$-R_{ab} \wedge e^a \wedge e^b + C_a \wedge C^a = d(C_a \wedge e^a).$$

The 4-form

$$\sqrt{-g}\,\varepsilon^{\mu\nu\lambda\rho}(R_{\mu\nu\lambda\rho} + \frac{1}{2}C^\alpha{}_{\mu\nu} C_{\alpha\lambda\rho}) \tag{33}$$

is thus seen to be the exterior derivative of the Chern-Simons type term:

$$-\sqrt{-g}\,\varepsilon^{\mu\nu\lambda\rho} C_{\nu\lambda\rho} \tag{34}$$

It follows from the identity (32) that

$$\int d^4x \sqrt{-g}\,\varepsilon^{\mu\nu\lambda\rho}(R_{\mu\nu\lambda\rho} + \frac{1}{2}C^\lambda{}_{\mu\nu} C_\alpha) = -\int d\sigma_\mu \sqrt{-g}\,\varepsilon^{\mu\nu\lambda\rho} C_{\nu\lambda\rho}. \tag{35}$$

Like the Euler class and the Pontryagin class, the left-hand side of the above equation is a topological invariant [12]. It is an invariant that characterizes the torsional topology of the underlying space-time; with vanishing torsion, each individual term on both sides of (32) and (35) vanishes.

The identity (32) can also be derived directly [8] from the Bianchi identity for non-vanishing torsion [11]. But, the derivation presented here has the advantage of making the meaning of the topological invariant more transparent. The geometric properties of the invariant have been studied by Chandia and Zanelli [12] and others [13]. It is well known that the Pontryagin or Chern-Weil class for the Lorentz group,

$$\int d^4x \sqrt{-g}\, \varepsilon^{\mu\nu\lambda\rho} R^{ab}{}_{\mu\nu} R_{ab\lambda\rho}, \tag{36}$$

is a topological invariant with an integral spectrum of values. Likewise, the Pontryagin class for the de Sitter or anti-de Sitter group,

$$\int d^4x \sqrt{-g}\, \varepsilon^{\mu\nu\lambda\rho} F^{AB}{}_{\mu\nu} F_{AB\mu\nu}, \tag{37}$$

is also a topological invariant with an integral spectrum of values. It is seen from (31) that the torsional topological invariant

$$\int d^4x \sqrt{-g}\, \varepsilon^{\mu\nu\lambda\rho} (R_{\mu\nu\lambda\rho} + \frac{1}{2} C^{\alpha}{}_{\mu\nu} C_{\alpha\lambda\rho}), \tag{38}$$

is proportional to the difference of the two Pontryagin classes (36) and (37), and thus has a discrete spectrum of values [12]. The dimensional constant $l$ appearing in (31), however, is an unknown parameter with no clear physical identification. In de Sitter type gravitational theories, the constant $l$ could conceivably play the role of the length characterizing the breakdown of the de Sitter symmetry.

Torsion is a geometric property not well investigated in mathematics. It is physicists' search for an extension of Einstein's general theory of relativity that torsion naturally appears. The properties of the torsional topological invariant (35) remain to be studied.